\documentclass[preprint,aps,showpacs,superscriptaddress,nofootinbib]{revtex4}

\usepackage[intlimits]{amsmath}
\usepackage{amssymb}
\usepackage{graphicx}
\usepackage{booktabs}
\usepackage{mathrsfs,slashed}
\usepackage{hyperref}
\usepackage{epsfig}
\usepackage{color}
\usepackage[super]{nth}

\begin{document}

\title{Surface waves in a collisional quark-gluon plasma}

\author{K. Baiseitov}
\affiliation{Institute for Experimental and Theoretical Physics,
Al-Farabi Kazakh National University, 71 Al-Farabi str., 050040 Almaty, Kazakhstan}

\author{Z. A. Moldabekov}
\affiliation{Institute for Experimental and Theoretical Physics,
Al-Farabi Kazakh National University, 71 Al-Farabi str., 050040 Almaty, Kazakhstan}

\author{D. Blaschke}
\affiliation{Bogoliubov  Laboratory  for  Theoretical  Physics,  Joint  Institute  for  Nuclear  Research,  141980  Dubna,  Russia}
\affiliation{Institute of Theoretical Physics, University of Wroclaw, Max Born pl. 9, 50-204 Wroclaw, Poland}
\affiliation{National Research Nuclear University (MEPhI), Kashirskoe Shosse 31, 115409 Moscow,  Russia}

\author{N. Djienbekov}
\affiliation{Institute for Experimental and Theoretical Physics,
Al-Farabi Kazakh National University, 71 Al-Farabi str., 050040 Almaty, Kazakhstan}

\author{T. S. Ramazanov}
\affiliation{Institute for Experimental and Theoretical Physics,
Al-Farabi Kazakh National University, 71 Al-Farabi str., 050040 Almaty, Kazakhstan}
\date{\today}

\begin{abstract}
Surface waves propagating in the semi-bounded collisional hot QCD medium (quark-gluon plasma) are considered. 
To investigate the effect of collisions as damping  and non-ideality factor, 
the longitudinal and transverse dielectric functions of the quark-gluon plasma are used within the 
Bhatnagar-Gross-Krook (BGK) approach. 
The results were obtained both analytically and numerically in the long wavelength limit. First of all, collisions lead to smaller values of surface wave frequency and their  stronger damping.
Secondly, the results show that non-ideality leads to the appearance of a new branch of surface waves compared to the collisionless case.  The relevance of the surface excitations (waves) for the QGP realized in experiments is discussed.   
\end{abstract}

\maketitle

\section{Introduction}

The quark-gluon plasma (QGP) is believed to be the state of matter in early universe, until $10^{-5}$ s after the Big Bang
the hadronization of the QGP sets in.
This fact motivated people to execute experimental research with ultra-relativistic heavy-ion collisions (HIC) at the 
super-proton synchrotron (SPS) and large hadron collider (LHC) of CERN Geneva and at the relativistic heavy-ion collider (RHIC) of the Brookhaven National Laboratory, where the temporary formation of a strongly coupled QGP state has been detected.

Quarks are building blocks of matter. Under ordinary conditions, they are strictly confined inside hadrons. 
Nevertheless, according to QCD as the fundamental gauge theory of strong interactions, quarks can be considered asymptotically free at high momenta (short distances). 
Such a state, when interactions among quarks and gluons can be considered sufficiently weak so that perturbation theory is
applicable can be achieved in deep inelastic scattering experiments.
However, when hadronic matter is compressed to high densities of about $\rho \sim 1 $ fm$^{-3}$ and heated to high temperatures, exceeding the pseudocritical temperature of $T_c = 156.5\pm 1.5$ MeV as it is determined in lattice QCD simulations \cite{Bazavov:2018mes} at vanishing baryon density, the QGP is in a strongly coupled plasma regime close to the hadronisation transition.
For this reason, it is legitimate to consider in this regime collective effects in the QGP as a many-body system of quarks and gluons.
Many-body processes in the strongly coupled QGP manifest themselves by the occurrence of collective modes, such as plasmons \cite{bonitz1994.1, bonitz1994.2}.
"Colored" plasmons were studied within a perturbative approach, because they were considered in a state well above the 
pseudocritical temperature. 

One may consider the collision rate in a QGP in generalization of approaches that were previously developed, and study their effect on observable processes.
For this aim, the dielectric functions for an ultra-relativistic plasma with collisions were analyzed 
on the basis of the Bhatnagar-Gross-Krook (BGK) collision integral \cite{BGK:1954} for the kinetic equation of the QGP \cite{Chakraborty:2007ug, carrington2004}. 
Furthermore, the wakefield generated by partons moving inside the QGP was investigated within a collisional approach \cite{Chakraborty:2007ug}, in analogy with wakes in QGP without collisions \cite{Chakraborty:2006md}. 
The wakefield has a different shape than in a static potential case, and the potential affects physical quantities, such as 
ground state, binding energy etc. 
The same approach to the collision rate was used to investigate plasmons of waves in the QGP with anisotropy \cite{carrington2014}.

From this review, it is clear that collective modes are particularly interesting for the QGP research. 
Most of works are done on waves propagating inside the matter with infinite boundaries and various models have been applied to describe them \cite{jiang2010, jiang2011}.
One of the first works that proposed to study the quenching of jets inside the QGP as an indicator for the creation of a deconfined state is Ref.~\cite{Gyulassy:1990ye}.

However, the plasma inevitably has to have a boundary when we consider the case of QGP creation in a HIC experiment.
Consequently, once there is a boundary of the plasma region, there shall be surface waves.
Of course, those surface waves have an impact on transport properties of the medium, i.e. they have to affect on quarks moving close to the surface of the medium.
Basically, all we can detect in the experiment are the particles that reach the detector.
Let us consider a test particle that first moves inside the medium (where it interacts with waves inside the medium), then it crosses through the boundary (where it interacts with surface waves) and finally it hits the detector.

Previous works focused mainly on in-medium waves, while surface waves were ignored. To this end, we present results for surface waves in this work.


\section{Formulation of the problem and results}

In this section we present the dispersion equation with relativistic velocities and dielectric functions as matter properties in 
order to find surface waves of the quark-gluon plasma state.
We develop the solution analytically with some limit and we determine additional points on the graph that was found numerically with limits only in machine precision.


\subsection{Dispersion equation}

Let us consider a semi-infinite quark-gluon plasma that occupies some region of space.
We call it semi-infinite as it is bounded only by one side and we restrict ourselves to a plane surface as the simplest case for a boundary condition.
The plasma resides at the positive values of $x$-axis ($x>0$) and borders the vacuum that persists on the opposite side of the $x$-axis ($x<0$).
This means that at $x=0$ we have flat surface and the problem has cylindrical symmetry because of this condition.
The mirror reflection of particles that fall onto this surface means that we can apply $\delta f(0, v_x > 0) = \delta f(0, v_x < 0)$ for the perturbed distribution function of plasma.
Therefore, the dispersion relation for surface waves in the plasma will be \cite{ruhadze1984}

\begin{equation}
    \label{dispersionRuhadze}
    \sqrt{\frac{k_z^2}{\omega^2} - 1} + \frac{2\omega}{\pi}
    \int^\infty_0 \frac{dk_x}{k^2}
    \left(\frac{k_z^2}{\omega^2 \epsilon_l} - 
    \frac{k_x^2}{k^2 - \omega^2\epsilon_{tr}}\right) = 0.
\end{equation}

In this article natural units are used, i.e. $\hbar = c = 1$. 
Also, we have a two dimensional problem, which means $k^2 = k_x^2 + k_z^2$.

Note that surface waves are well studied for an ordinary electromagnetic plasma and its dispersion relations are well known \cite{jung2006, jung2008, jung2013}. 
Those articles studied quantum effects in the electromagnetic plasma. 
We use the same approach and investigate oscillation modes of the quark-gluon plasma.

As it was said in the Introduction, the BGK approach can describe dielectric functions for a collisional QGP and the longitudinal and transverse dielectric functions are described following \cite{Chakraborty:2007ug}

\begin{equation}
    \label{epsilonLong}
    \epsilon_{l}(\omega, k)=1+\frac{m_{D}^{2}}{k^{2}}\left(1-\frac{\omega+i \nu}{2 k} \ln \frac{\omega+i \nu+k}{\omega+i \nu-k}\right)\left(1-\frac{i \nu}{2 k} \ln \frac{\omega+i \nu+k}{\omega+i \nu-k}\right)^{-1},
\end{equation}

\begin{equation}
    \label{epsilonTran}
    \epsilon_{t}(\omega, k)=1-\frac{m_{D}^{2}}{2 \omega(\omega+i \nu)}\left\{1+\left[\frac{(\omega+i \nu)^{2}}{k^{2}}-1\right]\left(1-\frac{\omega+i \nu}{2 k} \ln \frac{\omega+i \nu+k}{\omega+i \nu-k}\right)\right\},
\end{equation}
where $\nu$ is the collision rate.  
The Debye screening mass can be written explicitly as $m_D = \sqrt{1 + N_f / 6}~gT$,
with $N_f$ being the number of quark flavours in the QGP and $g$ the coupling constant of the strong interaction.
The solution can be given in dimensionless form by scaling all dimenionful quantities with the Debye screening mass. 
Therefore, the number of quark flavours does not influence on the structure of the result and can be arbitrary. 
This can be used in the calculation that is discussed in the Conclusion.


\subsection{Long wavelength limit}

The dispersion relation \eqref{dispersionRuhadze} represents a non-linear integral equation and, in general,
it is impossible to find the exact solutions of this equation analytically. 
When considering the long wavelength limit, for our case we assume $|k/(\omega + i \nu)| \ll 1$.
One needs to take into account that $\omega$ has real and imaginary part.
The imaginary part of $\omega$ refers to either damping or growth of the amplitude of the wave.
In our case, there has to be only a damping of the wave since we have no external source of energy and the existence of collisions dictates that we have interaction between the wave and the particle in a medium that takes the energy away 
from the waves, which is the so called dissipation effect.

For the following numerical and analytical calculations we make the quantities dimensionless (as mentioned above) 
by using the Debye mass $m_D$ 
\begin{equation}
    \label{dimensionless}
    \widetilde{\omega} = \frac{\omega}{m_D};\;
    \widetilde k =\frac{k}{m_D};\;
    \widetilde\nu = \frac{\nu}{m_D}.
\end{equation}

Bearing in mind that we introduced the small parameter as $a = |k/(\omega + i \nu)| \ll 1$, Eqs. \eqref{epsilonLong} and \eqref{epsilonTran} yield 
\begin{eqnarray}
    \epsilon_l(\widetilde \omega, \widetilde k) &=& 1 + \frac{1}{\widetilde k^2}
    \left[
        1 - \frac{1}{2a}\ln{\left(\frac{1 + a}{1 - a}\right)}
    \right]
    \left[
        1 - \frac{i\widetilde \nu}{2\widetilde k}\ln{\left(\frac{1 + a}{1 - a}\right)}
    \right]^{-1} \nonumber\\
    &\approx&
    1 + \frac{1}{\widetilde k^2}
    \left(
        -\frac{a^2}{3} - \frac{i\widetilde \nu a^3}{3\widetilde k}
    \right) \approx 
    1 - \frac{1}{3(\widetilde \omega +i\widetilde\nu)^2},
\end{eqnarray}
and
\begin{eqnarray}
    \epsilon_{tr}(\widetilde \omega, \widetilde k) &=& 1 - \frac{a}{2\widetilde \omega \widetilde  k}
    \left[
        1 + \left( \frac{1}{a^2} - 1\right)
        \left(
            1 - \frac{1}{2a}\ln{\left(\frac{1 + a}{1 - a}\right)}
        \right)
    \right]  \nonumber\\
    &\approx& 
    1 - \frac{a}{2\widetilde  \omega \widetilde  k} \left(
        \frac{2}{3} + \frac{2 a^2}{15}
    \right)
    \approx
    1 - \frac{1}{3\widetilde  \omega(\widetilde  \omega + i\widetilde  \nu)}.
\end{eqnarray}
Summarizing these calculations we present the results for the long wavelength limit again
\begin{eqnarray}
    \label{epsilonLongSeries}
     \epsilon_l( \omega) &=& 1 - \frac{m_D^2}{3~(\omega + i \nu)^2},\\
    \epsilon_{tr}( \omega) &=& 1 -  \frac{m_D^2}{3~\omega( \omega + i  \nu)}.
    \label{epsilonTranSeries}
\end{eqnarray}
{The intermediate steps of the derivation are given in the Appendix.}

In the next step, we substitute the expressions \eqref{epsilonLongSeries} and
\eqref{epsilonTranSeries} into the dispersion equation \eqref{dispersionRuhadze} and use $\omega=\widetilde{\omega}m_D$. 
Finally, after integration and algebraic operations we get
\begin{equation}
    \label{dispersionSeries}
    \sqrt{\frac{k_z^2}{\omega^2} - 1} + 
    \frac{k_z}{\omega \left(
        1 - \frac{m_D^2}{3(\omega + i \nu)^2}
        \right)} + 
    \frac{\sqrt{k_z^2 - \omega^2 \left(
        1 - \frac{m_D^2}{3\omega(\omega + i \nu )}
        \right) } - k_z}
    {\omega \left(
        1 - \frac{m_D^2}{3\omega (\omega + i \nu)}
        \right) } = 0
\end{equation}
This is a non-linear algebraic equation of \nth{4} order, and it has solutions for complex $\omega$. 

\begin{figure}[t]
    \centering
    \includegraphics[width = 16cm]{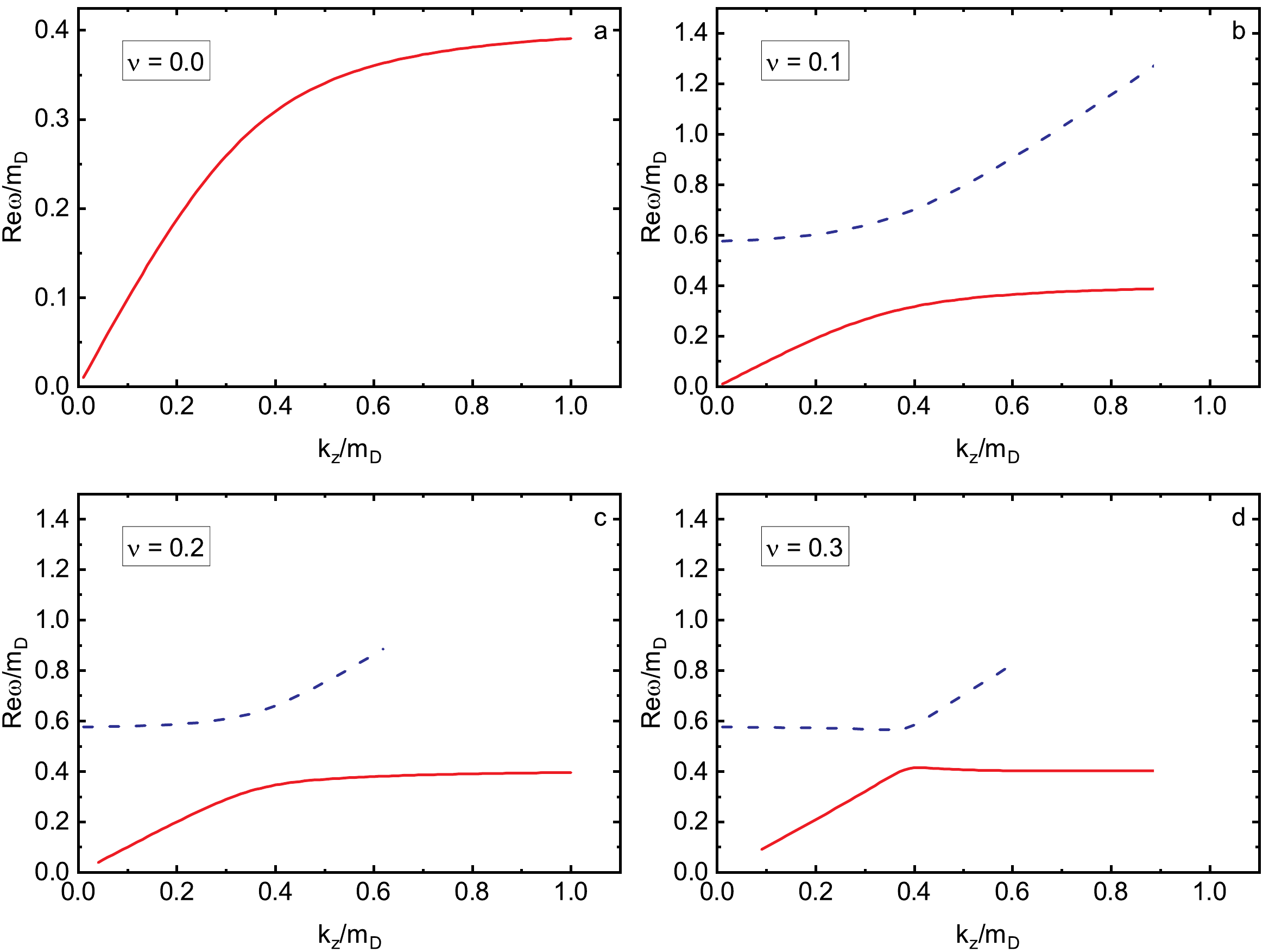}
    \caption{Real part of the dispersion $\omega(k)$ of surface waves for different collision rates $\nu$. Solid and dashed lines correspond to two different modes of surface oscillations. The mode corresponding to the dashed line appears only in the collisional case with $\nu \neq 0$.}
    \label{realFig}
\end{figure}

\begin{figure}[h]
    \centering
    \includegraphics[width = 16cm]{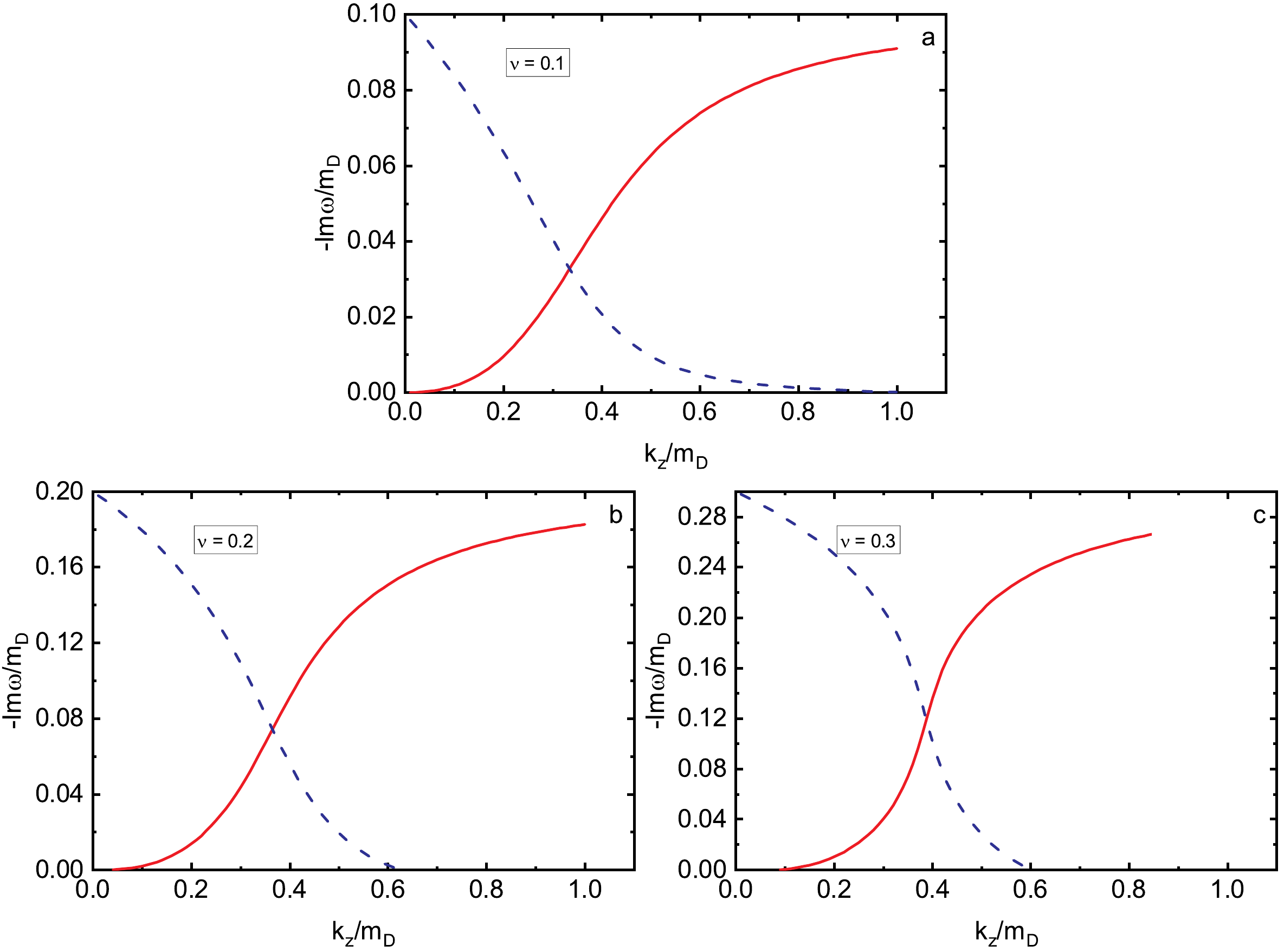}
    \caption{Imaginary part of the dispersion $\omega(k)$ of surface waves for different collision rates $\nu$; note that $\omega$ has no imaginary part for $\nu = 0$. Solid and dashed lines correspond to two different modes of surface oscillations. The mode corresponding to the dashed line appears only in the collisional case with $\nu \neq 0$.}
    \label{imagFig}
\end{figure}

The real part of the solution of equation  \eqref{dispersionSeries} shown in Fig.~\ref{realFig} corresponds to the dispersion modes and the imaginary part shown in 
Fig.~\ref{imagFig} refers to damping.
However, depending on the collision rate $\nu$ we get a different number of roots for this equation and we obtained also a negative dispersion mode.
Since the negative dispersion is considered to be unphysical it has not been plotted.

In the collisionless regime, when $\nu = 0$, there is only one dispersion mode (see Fig.~\ref{realFig}), and the imaginary 
part of this mode is zero. 
For this reason it is not shown in the second figure (see Fig.~\ref{imagFig}).
The inclusion of collisions gives a new dispersion mode that is absent in the collisionless case.
Remarkably, this new mode due to a finite collision rate does not appear for surface waves in an ordinary electromagnetic plasma.

One of the modes (the solid line in Fig.~\ref{realFig}) is weakly damped at small wave numbers. 
It saturates in frequency with increasing wave-number and its damping coefficient increases as well 
(see the solid line in Fig.~\ref{imagFig}). 
At $\nu/m_D=0.2$ and $\nu/m_D=0.3$, this mode does not appear at $k_z/m_D<0.05$ and at $k_z/m_D<0.1$, respectively. 

At $\nu/m_D=0.1$, the second mode (dashed line in Fig.~\ref{realFig} b) lies well above the light line $\omega = k_z$ and shows weak or no damping at  $k_z/m_D>0.6$ (see dashed line in Fig.~\ref{imagFig} a).
The second mode at $\nu/m_D=0.2$ and $\nu/m_D=0.3$ (dashed line in Fig. \ref{realFig} c and d) exists only  up to $k_z/m_D=0.6$ 
and is strongly damped (see dashed line in Fig.~\ref{imagFig} b and c).


\subsection{Relevance to the experimental QGP parameters}

Clearly, in experiments, the finite volume of the QGP defines the realizable wavenumber values.  
Therefore, we provide evaluation of the wavenumber range relevant to the experimental QGP.  
To begin with, for a deconfined phase in which QGP, the renormalized coupling constant should be $\alpha_s<1$. 
For the corresponding constant of strong interaction, evaluated using $g=\sqrt{4\pi \alpha_s}$, we have $g<3.54$. 
Next, computing the screening mass as $m_D=g T\sqrt{1+N_f/6}$, we find $m_D<678.52~(715.22)~{\rm MeV}$ for temperature $T=156.5 ~{\rm MeV}$ and $N_f=3~(4)$ quarks flavors. 
The size of a spherical fireball with diameter $d$ is in the range from  $d=2.56~{\rm fm}$ to $d=18.84~{\rm fm}$, see \cite{PhysRevC.73.044905}. 
Using the value of $m_D$, we find that the dimensionless wavenumber is $k/m_D\gtrsim 0.1$  for $d=2.56~{\rm fm}$ and $k/m_D\gtrsim 0.015$ for $d=18.84~{\rm fm}$. Obviously, obtained wavenumber range extends towards smaller values of $k/m_D$ at higher temperatures. Therefore, the long wavelength limit is expected to be relevant for the discussion of the QGP in high-energy heavy-ion collision experiments.


\section{Conclusion}

We have considered surface waves of a quark-gluon plasma, because every heavy-ion collision experiment creating it has a finite size (as discussed in Sec. II.C). 
Therefore, in theoretical investigations the fireball and the plasma in it should in principle be treated as an inhomogeneous system so that it is interesting to study whether such conditions are likely to change the expected results of the experiments.

The appearance of a new oscillation mode with the inclusion of a nonzero collision rate is a phenomenon of particular interest, which is not present for a classical plasma.
Moreover, after analytical treatment the dispersion equation seems to be solvable. 
However, as one can see from the graphs, the dispersion modes do not exist for some values of the wave number $k_z$. 
For this reason, the nature of these oscillations has to be analyzed more fundamentally and raises particular interest.

Note that it was possible to execute the calculation in the long wavelength limit only and the reader needs to understand the restriction that is imposed to the solution. However, for applications to the quark-gluon plasma created in heavy-ion collisions, due to the finite size of such a system, the exact solution of the Eq.~(\ref{dispersionSeries}) must be found using the dielectric function without the long wavelength approximation, while in this work we present as a first step results obtained using the dynamic dielectric function in the long wavelength limit. 
The authors plan to continue the numerical evaluation to be reported in forthcoming publications.

As a final remark, we note that it is of interest to study {\it first} more in detail the surface waves in viscous and strongly correlated regimes using a proper dielectric function (e.g. see Refs. \cite{jiang2010, jiang2011, PeraltaRamos:2012er}), 
and to investigate in a {\it second} step the interaction of produced particles with the surface waves  in ultra-relativistic 
heavy-ion collision experiments.  

\section*{Acknowledgments}
This work has been supported by the MES RK under Grant No. BR05236730 (2020) and by RFBR under grant number 
18-02-40137.

\appendix

\section{Appendix}

Taking into account that $a\ll1$, we used the following expansion:
\begin{equation}
\label{eq:a1}
        \ln \frac{1 + a}{1 - a} \approx 2a + \frac{2a^3}{3} + \frac{2a^5}{5} + \dots
\end{equation}
Substituting the expansion (\ref{eq:a1}) into Eqs. (5) and (6) we get


\begin{equation}
\label{eq:a2}
    \epsilon_{l}(\widetilde \omega, \widetilde k) \approx 
    1+\frac{1}{\widetilde k^{2}} \left(
    - \frac{a^2}{3} - \frac{a^4}{5}
    \right) 
    \left(
    1-\frac{i \widetilde\nu}{\widetilde k} a - \frac{i \widetilde\nu}{3 \widetilde k} a^3 - \frac{i \widetilde\nu}{5 \widetilde k} a^5
    \right)^{-1},
\end{equation}
and 
\begin{equation}
\label{eq:a3}
    \epsilon_{t}(\widetilde \omega,\widetilde k) \approx
    1 - \frac{a}{2 \widetilde \omega \widetilde k} \left\{
    1 + \left(
    \frac{1}{a^2} - 1
    \right) 
    \left(
    - \frac{a^2}{3} - \frac{a^4}{5}
    \right) 
    \right\},
\end{equation}
respectively.
Neglecting terms of $\mathcal{O}(a^3)$ in Eqs.~(\ref{eq:a2}) and (\ref{eq:a3}), and making use of the expansion 
$ \left(1 - \frac{i \widetilde \nu}{\widetilde k} a\right)^{-1} \approx 1 + \frac{i \widetilde \nu}{\widetilde k}a$ in Eq.~(\ref{eq:a2}), we arrive at
\begin{equation}
\label{eq:a4}
    \epsilon_{l}(\widetilde \omega,\widetilde k) \approx 
    1+\frac{1}{\widetilde k^{2}} \left(
    - \frac{a^2}{3}
    \right) 
    \left(
    1-\frac{i \widetilde\nu}{\widetilde k} a
    \right)^{-1} \approx 
    1+\frac{1}{\widetilde k^{2}} \left(
    - \frac{a^2}{3}
    \right) 
    \left(
    1 + \frac{i \widetilde \nu}{\widetilde k} a
    \right),
\end{equation}
and
\begin{equation}
\label{eq:a5}
    \epsilon_{t}(\widetilde \omega, \widetilde k) \approx
    1 - \frac{a}{2 \widetilde \omega \widetilde k} \left\{
    1 + \left(
    \frac{1}{a^2} - 1
    \right) 
    \left(
    - \frac{a^2}{3} - \frac{a^4}{5}
    \right) 
    \right\} = 
    1 - \frac{a}{2 \widetilde \omega \widetilde k} \left\{
    \frac{2}{3} + \frac{2 a^2}{15} 
    \right\},
\end{equation}
Finally, dropping terms of $\mathcal{O}(a^3)$ in Eqs.~(\ref{eq:a4}) and (\ref{eq:a5}),  we get expressions \eqref{epsilonLongSeries} and \eqref{epsilonTranSeries}.



\end{document}